\documentclass[a4paper,11pt]{article}
\usepackage{pos}
\usepackage{braket}
\usepackage[capitalize]{cleveref}

\title{The electromagnetic pion form factor and its phase}

\author*[a]{Pablo Sanchez-Puertas}
\author[a,b]{Enrique Ruiz Arriola}

\affiliation[a]{Departamento de Física Atómica, Molecular y Nuclear, Universidad de Granada,\\
Av. de la Fuente Nueva s/n, E-18071, Granada}

\affiliation[b]{Instituto Carlos I de Física Teórica y Computacional, Universidad de Granada,\\
Av. de la Fuente Nueva s/n, E-18071, Granada}

\emailAdd{pablosanchez@ugr.es}
\emailAdd{earriola@ugr.es}

\abstract{
In this work, we employ a dispersion relation that allows to recover the electromagnetic 
form factor of the pion from its modulus along the unitarity cut. Unlike the 
standard dispersive approach, this method extends in a straightforward manner above the 
inelastic region, allowing us to obtain the form factor's phase, as well as its real and 
imaginary parts. As applications, we extract the $P$-wave $\pi\pi$ phase shift, the 
octet form factor, derive bounds in the spacelike region, and offer insights into the 
onset of perturbative QCD.  
}

\FullConference{10th International Conference on Quarks and Nuclear Physics (QNP2024)\\
8-12 July, 2024\\
Barcelona, Spain\\}

\begin{document}
\maketitle

\section{Introduction}

The electromagnetic form factor of the pion, $F_Q^{\pi}(q^2)$, is defined in terms 
of the matrix element
\begin{equation}
  \bra{\pi^+(p')} J_Q^{\mu}(0) \ket{\pi^+(p)}  = F_Q^{\pi}(q^2) (p+p')^{\mu}, \qquad 
  J_Q^{\mu} = \sum_q Q_q\bar{q}\gamma^{\mu}q, 
\end{equation}
where $q=p'-p$ is the momentum transfer. 
Such object encodes information about the pion internal structure 
and QCD dynamics. As a consequence, 
it inherits the non-perturbative nature of QCD, and few information is 
available from first principles besides its normalization, $F_Q^{\pi}(0)=1$,
and its high-energy behavior predicted by perturbative QCD 
(pQCD) \cite{Chen:2023byr}\footnote{We update the NLO expression in \cite{RuizArriola:2024gwb} 
with the NNLO calculation~\cite{Chen:2023byr}, adopting $\mu_R^2 \sim t $ and an asymptotic distribution amplitudes.}
\begin{equation}\label{eq:pQCD}
F_Q^{\pi}(-t) \to \frac{16\pi F_{\pi}^2 \alpha_s}{t} \left(1 + 6.58\frac{\alpha_s}{\pi} \left[ 1 + 11.75\frac{\alpha_s}{\pi} \right] \right), 
\end{equation}
yet the scale where this applies is not clear at all. 
This form factor plays an important role in hadronic physics at low and intermediate energies. 
For instance, it plays a central role in the muon (g-2)~\cite{Aoyama:2020ynm}, where its modulus, 
accessible via the $e^+e^-\to\pi^+\pi^-$ reaction, determines the $70\%$ of the 
full hadronic contributions. This has motivated a wealth of high-precision 
measurements that we will take advantage of in this work. 
Additional important information about the form factor is encoded 
in its phase, which, unlike the modulus, cannot be directly measured 
in experiments.
Of particular relevance is the phase of the form factor below 
the onset of inelasticities, which, aside from isospin-breaking 
(IB) corrections, is linked to the $P$-wave phase in elastic 
$\pi\pi$ scattering via Watson's theorem. 
The latter plays a key role in dispersive approaches to 
low-energy hadronic physics. 
Furthermore, the form factor’s phase beyond the inelastic 
threshold provides valuable insights into the onset of pQCD.
In our work~\cite{RuizArriola:2024gwb}, we extract such a phase 
using dispersion relations (DRs) that relate the phase of the 
form factor to its modulus above the unitarity cut. This contrasts with 
standard dispersive approaches building on the phase, that are not well-tailored to deal with inelasticities. 
Different applications are discussed: the extraction of $\delta_1^1$ and the octet form factor;
the extrapolation to the spacelike region; the onset of pQCD.

\section{Dispersion relation for the modulus: phase, real and imaginary parts \label{sec:PhaseReIm}}

The standard dispersive approach to reconstruct a variety of form factors 
demands partial knowledge of the relevant phase at hand. 
For the case of $F_Q^{\pi}$ this 
can be achieved, in the absence of zeros (see ~\cite{RuizArriola:2024gwb} 
for discussions on zeros), by writing down Cauchy's theorem 
for $\ln |F_Q^{\pi}(s)|$. This leads to the well-known
solution  that we will 
refer to as the phase-DR,
\begin{equation}\label{eq:OmnesSubt}
    F_Q^{\pi}(s) = \operatorname{exp}\left( 
    \frac{s}{\pi} \int_{s_{th}}^{\infty} dz \frac{\delta_Q^{\pi}(z) }{z(z-s)}
    \right) =
    \operatorname{exp}\left( 
    \frac{s}{\pi} \int_{s_{th}}^{s_{in}} dz \frac{\delta_Q^{\pi}(z) }{z(z-s)}
    \right)
    \operatorname{exp}\left( 
    \frac{s}{\pi} \int_{s_{in}}^{\infty} dz \frac{\delta_Q^{\pi}(z) }{z(z-s)}
    \right) .
\end{equation}
In general, $\delta_Q^{\pi}$ is unknown. However, below inelastic 
thresholds and up to isospin corrections that we discuss later, it can be 
identified with $\delta_1^1$, and the first integral on the right-hand 
side of \cref{eq:OmnesSubt}, known as the Omn\`es function, 
can be evaluated. The remaining integral for the inelastic region is 
phenomenologically parametrized as a series expansion.
Unfortunately, this approach cannot be applied for $s \gtrsim s_{in}$, 
where the phase is generally unknown and would demand a complicated 
coupled-channel analysis. 
To overcome this problem, in this work we will focus on a different 
dispersive approach based on Refs.~\cite{Truong:1968,Truong:1969gr,Geshkenbein:1969bb,Cronstrom:1973wr,Geshkenbein:1998gu}. 
This can be obtained, in the absence of zeros, by 
applying Cauchy's theorem to $\ln |F_Q^{\pi}(z)|/\sqrt{s_{th} -z}$ 
or $\ln |F_Q^{\pi}(z)/F_Q^{\pi}(s_{th})|/(s_{th} -z)^{3/2}$, leading 
to~\cite{RuizArriola:2024gwb}
\begin{align}
    F_{Q}^{\pi}(s) &{}= \operatorname{exp}\Big(
    \frac{s\sqrt{s_{th} -s}}{\pi}\int_{s_{th}}^{\infty}dz 
    \frac{\ln|F_Q^{\pi}(z)|}{z\sqrt{z -s_{th}}(z-s)}
    \Big), \label{eq:SubtZero} \\
    F_Q^{\pi}(s)  &{}= 
    F_Q^{\pi}(s_{th})^{1 - \left(\frac{s_{th} -s}{s_{th}}\right)^{3/2} } \operatorname{exp} \Bigg[
    -\frac{s(s_{th} -s)^{3/2}}{\pi}  \int_{s_{th}}^{\infty}dz 
    \frac{\ln|F_Q^{\pi}(z)/F_Q^{\pi}(s_{th})|}{z(z -s_{th})^{3/2}(z-s)}
    \Bigg], \label{eq:SubtThZero}
\end{align}
that we will refer to as DR1 and DR2. 
The relations above are remarkable: they provide a direct link 
among a measurable quantity, $|F_Q^{\pi}(s)|$ along the unitarity 
cut, and its value in the complex plane. 
Compared to the standard phase-DR, the modulus-DR opens up the 
possibility to unveil the inelastic region provided data for its 
modulus is available. 
In particular, the phase is given as
\begin{align}
    \delta(s) &{}=
    -\frac{s\sqrt{s -s_{th}}}{\pi} \operatorname{PV}\int_{s_{th}}^{\infty}dz 
    \frac{\ln|F_Q^{\pi}(z)|}{z\sqrt{z -s_{th}}(z-s)},\label{eq:SubtZeroPhase} \\
    \delta(s) &{}= 
    -\ln  F_Q^{\pi}(s_{th}) \left(\frac{s -s_{th}}{s_{th}}\right)^{3/2}
    -\frac{s(s -s_{th})^{3/2}}{\pi}  \operatorname{PV}\int_{s_{th}}^{\infty}dz 
    \frac{\ln|F_Q^{\pi}(z)/F_Q^{\pi}(s_{th})|}{z(z -s_{th})^{3/2}(z-s)},  \label{eq:SubtThZeroPhase}
\end{align}
that we will employ below in order to disentangle the real and 
imaginary parts of the form factor along the unitarity cut. 
Finally, an important sum rule can be derived from the DRs 
above:
\begin{equation}
    \frac{s_{th}^{3/2}}{\pi \ln F_Q^{\pi}(s_{th})} 
    \int_{s_{th}}^{\infty}dz 
    \frac{\ln|F_Q^{\pi}(z)/F_Q^{\pi}(s_{th})|}{z(z -s_{th})^{3/2}} =1, \label{eq:TheSumRule}
\end{equation}
that is key for consistency checks in our work~\cite{RuizArriola:2024gwb}. 

\begin{figure}
  \includegraphics[width=0.48\textwidth]{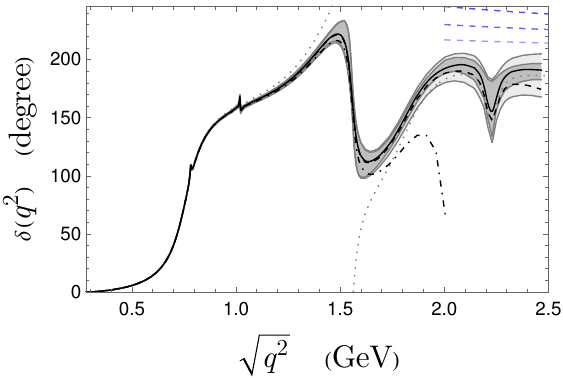}
  \includegraphics[width=0.48\textwidth]{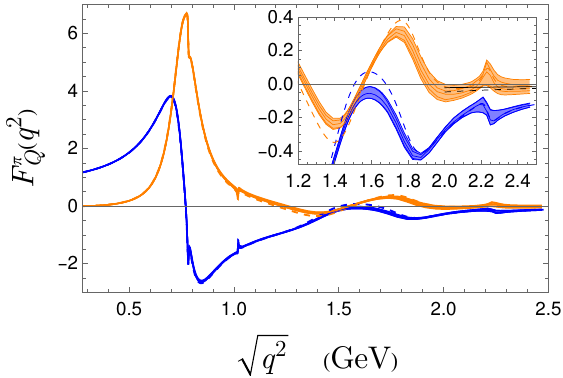}
  \caption{Left: the phase of the form factor (black line) with statistic(+systematic) 
  inner(outer) error bands. The dot-dashed/dashed black lines 
  are obtained for lower cutoffs in \cref{eq:SubtThZeroPhase}. The dotted 
  gray line is the original model phase. The dashed blue 
  lines, from light to dark, are the LO, NLO and NNLO pQCD prediction. 
  Right: The real(imaginary) parts as blue(orange) lines with corresponding error
  bads similar to the left figure. The dashed blue(orange)  
  lines are again the result from the model. The 
  solid(dashed) black lines are the NNLO pQCD prediction for the real(imaginary) part.   
  \label{fig:tl}}
\end{figure}

In the following, we make use of DR2 together with \textit{\textsc{BaBar}} data~\cite{BaBar:2012bdw} to 
extract the phase of the form factor, that can be achieved reliably up to around $2.5$~GeV. 
Remarkably, \textit{\textsc{BaBar}}'s data span energies 
from threshold up to $3$~GeV, and is amongst the most precise experiments, 
with subpercent uncertainties close to the $\rho$ peak. 
The reason to exclude other experiments are (i) the tensions with other 
experiments with similar precision, making impossible to combine them and 
(ii) the fact that it is the only precise experiment including data beyond 
the 1~GeV region, which is necessary to check the SR in \cref{eq:TheSumRule} 
and guaranteeing consistency~\cite{RuizArriola:2024gwb}. 
In order to perform the numerical integral, we interpolate the data with 
the help of a Gounaris-Sakurai model. 
Note that complex phases in the model break unitarity and analytic constraints, 
whereas the performed DR will restore the correct analytic properties shifting 
the phase appropriately.  
To keep track of correlations, we make use of a Monte Carlo (MC) analysis, 
following the method in Ref.~\cite{Ball:2009qv} to avoid d'Agostini bias. 
Further, we make sure that the sum rule in \cref{eq:TheSumRule} 
is fulfilled for every MC fit, that ensures consistency and is only possible 
with \textit{\textsc{BaBar}}'s high-energy points.
Our results for the phase, real and imaginary parts of the 
form factor are shown in \cref{fig:tl}, where the inner error bands 
represent the statistical uncertainty.  
The outer error band includes systematic uncertainties. These arise from 
the modelling below $0.6$~GeV; above this energy, the model yields results 
similar to a linear interpolation ---find details about this and the role of 
zeros in~\cite{RuizArriola:2024gwb}. 
As it can be appreciated, the phase approaches $\pi$ from above, as predicted 
by pQCD, yet the full result lies far from \cref{eq:pQCD}. 
A final remark is pertinent here and concerns the argument theorem in 
complex analysis. In our case of application, it implies that the 
asymptotic phase $\delta_Q^{\pi} \to \pi(1 + N-P)$~\cite{Truong:1968,RuizArriola:2024gwb}, where 
the first term comes from pQCD and $N(P)$ stand for the number of zeros(poles) in the 
first Riemann sheet. For complex-conjugate poles, $\delta_Q^{\pi} \to \pi(1 + 2n)$, 
but $n$ cannot be fixed from first principles (find comments on zeros in \cite{RuizArriola:2024gwb}).

\section{The $P$-wave $\pi\pi$ phase shift and isovector/isosinglet form factors}

\begin{figure}
  \includegraphics[width=0.48\textwidth]{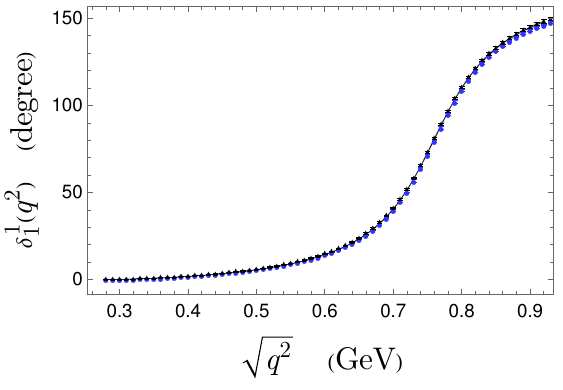}
  \includegraphics[width=0.48\textwidth]{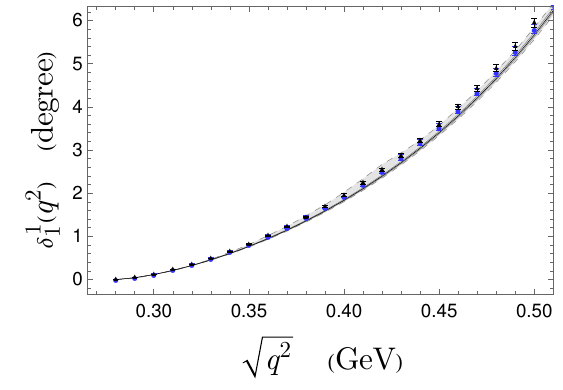}
  \caption{Left: the $\delta_1^1$ phase shift in the full elastic 
  region (black) compared to the result from \cite{Garcia-Martin:2011iqs}
  (blue bars) and \cite{Colangelo:2018mtw} (black bars). 
  Right: a detailed view of the low-energy region. 
  \label{fig:phaseshift}}
\end{figure}

As previously mentioned, $\delta_Q^{\pi}$ is related in the elastic 
region to $\delta_1^1$ modulo isospin corrections. 
To show how, we decompose the electromagnetic 
current as $J_Q^{\mu} = J_3^{\mu} +J_8^{\mu}/\sqrt{3}$, 
with $J^{\mu}_a=\bar{q}\gamma^{\mu}\frac{\lambda^a}{2} q$ and $\lambda^{a}$ 
Gell-Mann matrices, inducing the isospin decomposition 
$F_Q^{\pi} = F_3^{\pi} +F_8^{\pi}/\sqrt{3}$. 
Unitarity effects in the isovector form factor $F_3^{\pi}$ first arise 
from $I=1$ $\pi\pi$ interactions and, thereby, its phase identifies 
with $\delta_1^1$ in the elastic region. 
This contrast with the octet form factor, $F_8^{\pi}$, that 
should vanish in the isospin limit and features both $2\pi$ and $3\pi$ 
unitarity effects at the same order in IB. These materialize in the phenomenon known as 
$\rho-\omega$ mixing, clearly visible in the data, that allows to perform 
the isospin decomposition with little model dependence~\cite{Sanchez-Puertas:2021eqj,RuizArriola:2024gwb}. 
From the isovector one, we extract $\delta_1^1$ in the 
elastic region, which is shown in \cref{fig:phaseshift}.  
This is a key quantity in low-energy hadronic physics, which most
precise determination has been obtained using 
Roy equations~\cite{Garcia-Martin:2011iqs,Colangelo:2018mtw}, that
are shown in \cref{fig:phaseshift}. We find an 
excellent agreement with Ref.~\cite{Colangelo:2018mtw}, 
that also uses modern input from $e^+e^-\to\pi^+\pi^-$ data. 
Furthermore our results are competitive 
in precision with Ref.~\cite{Colangelo:2018mtw}, except 
at the low energy region (cf. \cref{fig:phaseshift} right)
where systematic effects dominate. 

\begin{figure}
  \includegraphics[width=0.48\textwidth]{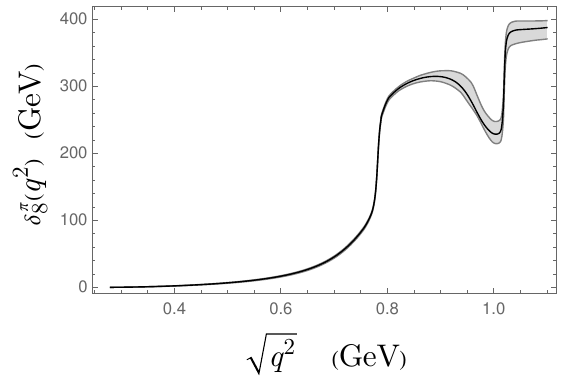}
  \includegraphics[width=0.48\textwidth]{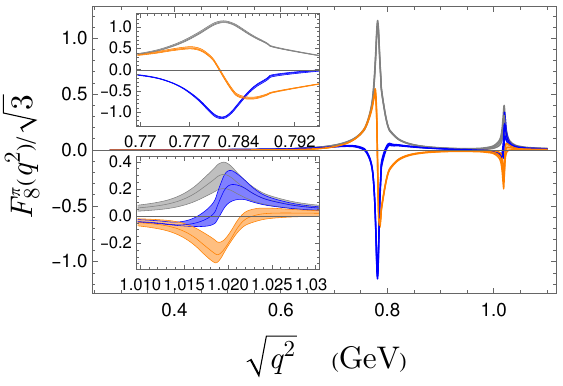}
  \caption{Left: the octet form factor phase (black) with 
  statistical error bands (gray).
  Right: The real (blue), imaginary (orange) and absolute value (gray) 
  of $F_8^{\pi}/\sqrt{3}$ with statistical error bands.  
  \label{fig:octet}}
\end{figure}

In addition, we provide our results for the octet form factor in 
\cref{fig:octet}. Note that neglecting strange-quark effects, this is 
related to the baryonic form factor discussed in 
Refs.~\cite{Sanchez-Puertas:2021eqj,Broniowski:2021awb}.
As it can be appreciated, the phase approaches 
$2\pi$ from above ---a value which is expected from 
the argument theorem and the real zero that 
follows from the null octet charge of the pion, $F_8^{\pi}(0)=0$. 

\section{Extrapolation to the spacelike region}

\begin{figure}
  \includegraphics[width=0.48\textwidth]{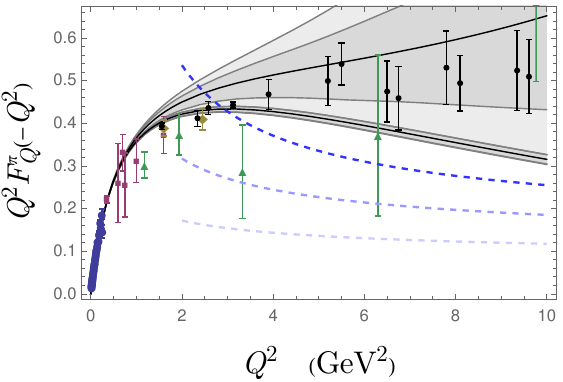}
  \includegraphics[width=0.5\textwidth]{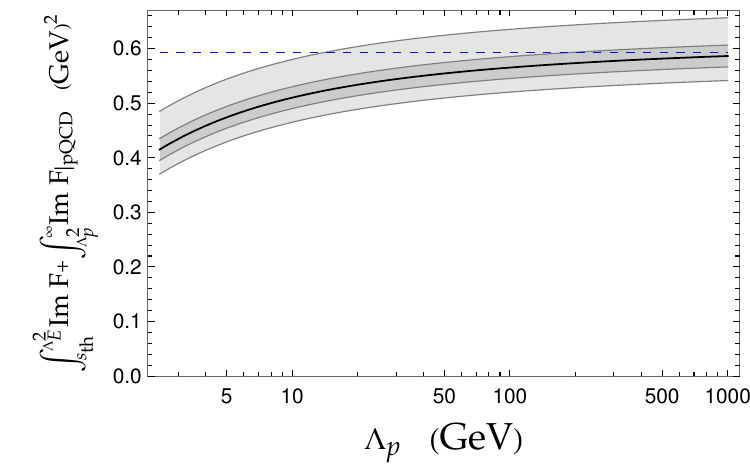}
  \caption{Left: Our upper and lower bounds for the spacelike 
  form factor (gray bands) compared to experimental data in colors (find 
  Refs. in \cite{RuizArriola:2024gwb}) and the new lattice 
  result~\cite{Ding:2024lfj} (black bars).
  Right: The SR2 in \cref{eq:SRs} for different choices of the 
  perturbative scale $\Lambda_p$.
  \label{fig:slsr}}
\end{figure}

The DRs in \cref{eq:SubtZero,eq:SubtThZero} can also be employed 
to obtain the value of the form factor in the spacelike region. 
Notoriously, provided the form factor lacks a steep and
sustained rise at large $q^2$ values,  
\cref{eq:SubtZero} and \cref{eq:SubtThZero} provide, 
respectively, lower and upper bounds~\cite{RuizArriola:2024gwb}. 
These are illustrated in \cref{fig:slsr} (left).
Interesting enough, the results resemble a VMD prediction (see also 
\cref{sec:SRpQCD}) and are far from pQCD, suggesting a late 
onset for the perturbative behavior. Since this is an attractive 
question itself, we discuss 
this property in the section below with the help of the sum rules.

\section{Sum rules and the onset of pQCD \label{sec:SRpQCD}}

Using Cauchy's theorem for $F_Q^{\pi}(s)$, the following sum rules 
can be derived (see Ref.~\cite{Donoghue:1996bt})
\begin{equation}
 \frac{1}{\pi} \int_{s_0}^{\infty} ds \frac{\operatorname{Im}F_Q^{\pi}(s)}{s} = 1 \quad (\textrm{SR1}),\quad
 \frac{1}{\pi} \int_{s_0}^{\infty} ds \operatorname{Im}F_Q^{\pi}(s) = 0 \quad (\textrm{SR2}), \label{eq:SRs}
\end{equation}
that follow from the normalization and pQCD asymptotics. 
These can be computed reliably from experiment using our result 
in \cref{sec:PhaseReIm} up to a cutoff $\Lambda_E=2.5$~GeV 
without the need of models. In this way, we obtain obtain 
$1.01(1)(^{+2}_{-1})$ for the SR1 (close to its nominal value), 
and $0.63(2)(^{+7}_{-4})~\textrm{GeV}^2$ for the SR2. 
Interesting enough, the latter is close to the simple VMD estimate, 
$M_{\rho}^2$, yet far from pQCD.  
One may wonder if the pQCD tail would explain such mismatch. 
To answer this question, we compute the pQCD contribution
by analytically continuing $\alpha_s$ to the timelike 
region\footnote{To do so, we take the LO result 
$\alpha_s = 4\pi/(\beta_0 L)$ and take 
$L = \ln(Q^2/\Lambda_{\textrm{QCD}}^2) \to \ln(s/\Lambda_{\textrm{QCD}}^2) -i\pi$.} 
and integrating from an unknown scale $\Lambda_p$ (where pQCD 
could hypothetically be applied) up to infinity. As we 
illustrate in \cref{fig:slsr} (right), this is not possible 
even if pQCD is applied already at $\Lambda_p = \Lambda_E$.
In consequence, this sum rule suggests sizeable contributions 
from the high-energy tail well above the pQCD prediction.

\section{Conclusions and outlook} 

Being the lightest hadron, the pion properties are a priori amongst the simplest
to be studied and measured. Yet, they continue posing challenges and attracting attention, 
particularly regarding its electromagnetic form factor. 
This is partly due to the role of the form factor in low-energy hadronic physics, where both, 
its modulus and phase, play an important role. 
Additionally, in the high-energy region, the applicability of pQCD and the 
transition from hadronic to quark-based descriptions remain an open and compelling questions. 
In this work, we have used a dispersive approach to extract both the modulus and phase, 
providing insight into these questions. 
In the future, we have plans to extend this approach to the Kaon case.

\acknowledgments

P.~Sanchez Puertas would like to thank Miguel {\'A}ngel Ojeda Garc{\'i}a for local 
hosting and hospitality during this workshop, as well as the organizers for the nice 
conference and atmosphere. This work has been supported by 
the Spanish Ministry of Science and Innovation under Grants No. PID2020–114767GB-I00 
and No. PID2023-147072NB-I00 by MICIU/AEI/10.13039/501100011033, as well as 
Junta de Andalucía under Grants No. POSTDOC\_21\_00136 and No. FQM-225.

\bibliographystyle{JHEP}
\bibliography{bibliography}

\end{document}